\documentclass[conference]{IEEEtran}

\pdfoutput=1

\usepackage{amsmath}
\usepackage{amsfonts}  
\usepackage{graphicx}
\usepackage[sort]{cite}

\title{Beam Switching Techniques for Millimeter Wave Vehicle to Infrastructure Communications}

\begin{document}

  \author{\IEEEauthorblockN{Hamed Mohammadi}
 \IEEEauthorblockA{Department of Electrical Engineering\\
 University of Kurdistan, Sanandaj, Iran.\\
 Email: Hamed.mohammadi@eng.uok.ac.ir}
 \and
 \IEEEauthorblockN{Reza Mohammadkhani}
 \IEEEauthorblockA{Department of Electrical Engineering\\
 University of Kurdistan, Sanandaj, Iran.\\
 Email: r.mohammadkhani@uok.ac.ir}}


\maketitle

\begin{abstract}
Beam alignment for millimeter wave (mm Wave) vehicular communications is challenging due to the high mobility of vehicles. Recent studies have proposed some beam switching techniques at Road Side Unit (RSU) for vehicle to infrastructure (V2I) communications, employing initial position and speed information of vehicles, that are sent through Dedicated Short Range Communications (DSRC) to the RSU. However, inaccuracies of the provided information lead to beam misalignment. Some beam design parameters are suggested in the literature to combat this effect. But how these parameters should be tuned? Here, we evaluate the effect of all these parameters, and propose a beam design efficiency metric to perform beam alignment in the presence of the estimation errors, and to improve the performance by choosing the right design parameters.
\end{abstract}
\quad

\begin{IEEEkeywords}
mm-wave; beam alignment; beam switching; vehicular communication
\end{IEEEkeywords}

\section{Introduction}
There is a worldwide interest and attempt to ease traffic congestion, in order to decrease traffic accidents and improve driving safety. As a real example of the motivation behind this, damage costs of traffic accidents in the EU countries is about 100 billion euro per annum  \cite{Heddebaut2010}. It can be reduced by employing communication technologies such as vehicle-to-vehicle (V2V), and vehicle-to-infrastructure (V2I) communications. These technologies allow each vehicle to share its radars and sensors data with other vehicles or the infrastructure\cite{González-Prelcic2016}.
However, current vehicles have an average number of 100 sensors per vehicle and it is expected to double by 2020, since vehicles are getting more intelligent and the Internet-of-things (IoT) is growing fast \cite{Choi2016}. This number of sensors and equipment, that can be used for safety applications \cite{Wang2016,Hobert2015}, and other potential applications such as  \cite{Va2016}: cooperative environmental perception, sensor data sharing, map downloading, sensor data gathering, and media downloading, needs a high data rate about multi giga-bits-per-second (Gbps). These applications can help to ease traffic on roads and improve safety (by providing 360-degree awareness, emergency break, awareness for entering the road at cross, and etc.) \cite{González-Prelcic2016,Choi2016,Uzcátegui2009}.

The current technology of vehicular communications, known as Dedicated Short Range Communications (DSRC), works at 5.9 GHz and it can only support a data rate of 6 Mbps to 27 Mbps \cite{V.Va2016}.

Available wideband channels at millimeter-wave (mm Wave) spectrum are good candidates to achieve such high data rates up to several Gbps \cite{Rappaport2013,Rangan2014,VaTransaction}.
It is also shown that mm Wave frequencies have less latency compared to the existing DSRC frequency bands  \cite{Cheng2011,Alejos2008}. However, mm Waves suffer from higher attenuation, which can be explained from the well-known Friis transmission equation given as follows. It defines the ratio of available power at the input of the receiver, $P_r$, to the output power from the transmitter, $P_t$, as:

\begin{equation}
\frac{P_r}{P_t}= G_t G_r \left(\frac{c_0}{4\pi f_c d}\right)^2
\end{equation}
where $G_t$ and $G_r$ are the antenna gains for the transmitter and the receiver, respectively, $c_0$ is the speed of light, $f_c$ is the carrier frequency, and $d$ is the distance between the transmitter and the receiver.  As we see, the received power decays by factors of $1/d^2$  and $\lambda^2=(c_0/f)^2$ , where $\lambda$  is the wavelength. As a result, mm Waves with lower wavelengths and higher frequencies, suffer from a higher loss in the received power (about 20 dB less than microwaves, with equal antenna gains at both cases), which needs to be compensated by increasing the antenna gains \cite{Molisch2016}. Array antennas are good candidates for mm Waves due to their narrow and directional beams and the ability to move beams electronically \cite{Komjani2006}. However, the transmitter and the receiver directional beams should be aligned, in order to achieve the maximum gain. Several beam alignment techniques are introduced in the literature \cite{González-Prelcic2016,Choi2016,Va2016,V.Va2016,Va2015,Mavromatis2017}.

Performing beam alignment in mm Wave vehicular communications is challenging due to the high mobility of vehicles. One solution in V2I communications, is to use DSRC data to have an initial estimate of speed and location of the vehicle, in order to reduce beam alignment overhead \cite{González-Prelcic2016}. It works as follows. When the vehicle enters the coverage area of a Road Side Unit (RSU), its speed and position information sends back to the RSU using DSRC. However, beam alignment correction might be needed to reduce the effect of an uncertain position estimation. In this method, each vehicle is always covered by a directional beam.

A similar work is reported in \cite{Va2015} for high speed trains, that uses the train control system to have an estimate of train position, and to achieve an efficient beam switching approach. Each RSU has a number of directional narrow beam, that switches to the right beam based on the obtained information and a prediction algorithm. Article \cite{V.Va2016} also uses this idea connecting vehicles to the beam switching based RSU in V2I scenarios. It suggests to have some overlap between beams to reduce the effect of speed estimation error. However, it does not take into account the positioning error, and a constant speed is assumed for each vehicle while passing the coverage area of the RSU. Furthermore, \cite{Mavromatis2017} proposes a similar beam switching based beamforming algorithm that performs better by combining the position, motion and velocity data of each vehicle.

In this paper, we evaluate the effect of beamwidth, or equivalently the number of beams per RSU, and overlapping between beams on the system performance. We consider two available beam design approaches: equal beam, and equal coverage area for all beams. We finally propose a new a beam design efficiency criterion that allows as to select the right values of design parameters and decide which beam design approach is better in the presence of estimation error.

The rest of this paper is organized as follows. We introduce the system model in Section II. Then, average data rate and average outage time are addressed in section III as the system performance metrics. It follows by introducing our proposed beam design efficiency metric that seeks a tradeoff between the average data rate and the average outage time. It is desired to increase data rate and decrease outage time. Finally, Section V concludes the paper.

\begin{figure}[t]
\centering
\includegraphics[width=0.99\linewidth]{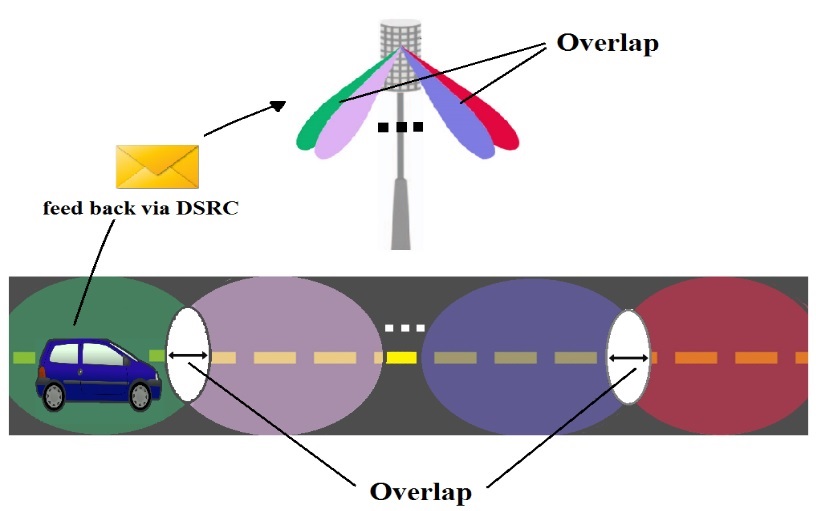}
\caption{System model for Beam alignment}
\label{fig:RSU_beam}
\end{figure}

\begin{figure}[h]
\centering
\includegraphics[width=0.99\linewidth]{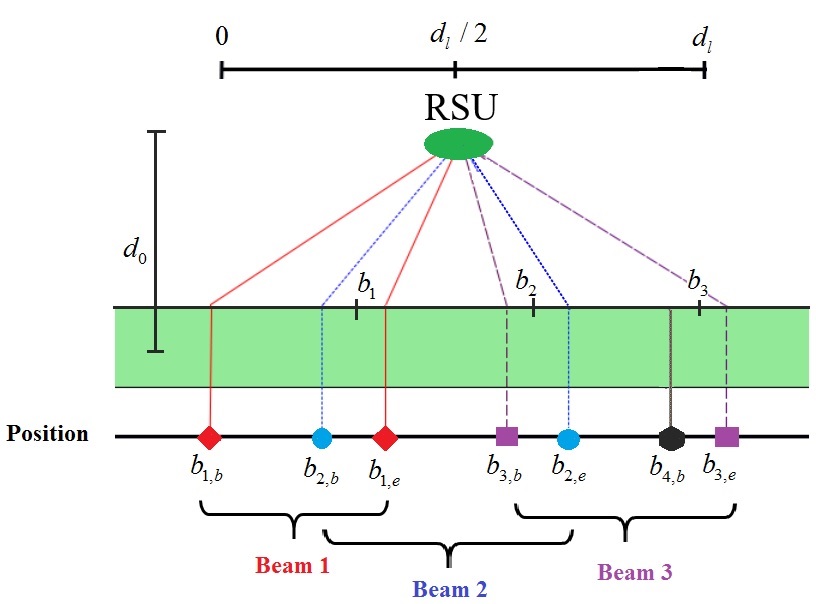}
\caption{Beam design considering overlap between beams}
\label{fig:RSU_beam_sectors}
\end{figure}

\section{System Model}\label{sec:model}
A system of two-lane highway covered by a number of RSUs is considered. We assume that vehicles have a constant speed of 90 km/h while passing coverage area of an RSU. (However, this speed value is an example and changing it does not affect the final result). We suppose RSUs are mounted on lighting poles, so line-of-sight components are most likely available.

When a vehicle enters the coverage area of the RSU, sends its initial position and speed information to the RSU thorough DSRC. Then, the RSU predicts next positions of the vehicle and assigns a beam until it passes the coverage area. However, in order to reduce beam alignment overhead, initial information of the vehicle is sent once and it might contain errors. In the case of large position error, beam alignment needs to be corrected, and the right beam selected. For the case of speed estimation error, we consider the vehicle speed estimate as:
\begin{equation}
\hat{v} = v + v_e
\end{equation}
where $v$  is the real and accurate speed of the vehicle, and  $v_e$  is the estimation error assumed to be Gaussian with zero-mean and $\sigma^2_v$   variance.

We assume each RSU has $N_b$   beams to cover its assigned area, and these beams can overlap to reduce the effect of speed estimation error. Two scenarios are employed for beam switching design: all beams have either i) equal beamwidth, or ii) equal coverage area.
\section{System Performance}
We consider average data rate and average outage time as the two performance metrics of the system.

\subsection{Average Data Rate}
Assuming analog beamforming at both sides of the link (RSU and vehicle), the received power is given by \cite{V.Va2016}:
\begin{equation}
P_r(t,\theta_b)=\frac{A G_r(\theta_b)}{\left[(vt-d_l/2)^2+d_{el}^2\right]^{n/2}}
\end{equation}
where  $\theta_b$ and $G_r$  are the beamwidth and antenna gain of the receiver respectively, $d_l$  is the distance covered by the RSU, and $n$  is the path loss exponent. And, $d^2_{el}$  can be calculated as follows:
\begin{equation}
d^2_{el}=d^2_0+(H_R-H_V)^2
\end{equation}
where  $H_R$ and $H_V$  are heights of the RSU and the vehicle, respectively, and $d_0$  is shown in Fig. \ref{fig:RSU_beam_sectors}. Parameter $A$ is also obtained from the following equation:
\begin{equation}
10\log(A) = \textit EIRP_{dBm}-w+10n \log_{10}(\lambda/4\pi)
\end{equation}
where  $w$ represents the shadowing margin, $\lambda$  is the signal wavelength, and $\textit EIRP_{dBm}$  is the effective isotropically radiated power. Assuming a beam with negligible sidelobes, an approximate of the receive antenna gain can be expressed as \cite{Balanis2005}:
\begin{equation}
G_r(\theta_b)\approx \frac{\pi^2}{\theta_{el}\theta_b}
\end{equation}
where  $\theta_b$ is the azimuth beamwidth and $\theta_{el}$  is the elevation beamwidth. We can obtain $\theta_{el}$  from
\begin{equation}
\theta_{el} = \tan^{-1}\left(\frac{d_0+2W_l}{H_R}\right) - \tan^{-1}\left(\frac{d_0}{H_R}\right)
\end{equation}
where  $W_l$ is the width of each lane in our model.

We also note that $v_e=\hat{v}-v$  can have either negative or positive values that should be involved in relations separately. As illustrated in Fig. \ref{fig:RSU_beam_sectors}, each interval $[b_{i,b},b_{i,e}]$  represents the coverage area of $i$-th  beam, and the switching is happened at the middle of   $[b_{i+1,b},b_{i,e}]$.

We can express the instantaneous channel capacity for a given bandwidth $B$  as
\begin{equation}
C(t,\theta_b) = B\log_2\left(1+\rho(t,\theta_b)\right)
\end{equation}
where $\rho(t,\theta_b)$ is the instantaneous signal-to-noise ratio (SNR) given by
\begin{equation}
\rho(t,\theta_b) = \frac{P_r(t,\theta_b)}{P_{noise}}
\end{equation}
and $P_{noise}$ is the noise power, characterized by
\begin{equation}
P_{noise} = N_{floor}+10\log_{10}(B)+NF
\end{equation}
where  $N_{floor}=-174\: dBm$ is the noise floor, and  $NF$ is the noise figure. The average data rate can be determined by integrating over $C(t,\theta_b)$   with respect to the time $t$  that each beam is aligned. Theoretically, switching to the  $i$th beam can occur at the time $b_i/v$, however $v$  is not known and it needs to be replaced by an estimate $\hat v$  of the speed. Therefore, the total amount of data received by the beam $i$  becomes \cite{V.Va2016}:

\begin{equation}
D_i(\theta|v_e\geq0)= B\int_{\max\{\frac{b_{i,b}}{v}, \frac{b_{i-1}}{v+v_e}\}}^{\frac{b_i}{v+v_e}} \log(1+\rho(t,\theta_i))dt
\end{equation}

\begin{equation}
D_i(\theta|v_e<0)= B\int_{ \frac{b_{i-1}}{v+v_e}}^{\min\{\frac{b_{i,e}}{v},\frac{b_i}{v+v_e}\}} \log(1+\rho(t,\theta_i))dt
\end{equation}
where $\theta_i$ is the beamwidth of the beam $i$ and $i=1,2,...,N_b$, having $N_b$ as the number of beams that cover whole area of each RSU. According to the boundaries depicted in Fig. \ref{fig:RSU_beam_sectors}, we get $v_e>\frac{b_{i}-b_{i,b}}{b_{i,b}}$ for non-negative values of $v_e$, and $v_e<\frac{b_{i-1}-b_{i,e}}{b_{i,e}}$  for $v_e<0$. If we have no overlap between beams, then $b_{i,b}=b_{i-1,e}=b_{i-1}$. Finally, we can average the data rate of $i$th  beam with respect to the random variable $v_e$ as follows:
\begin{align}
\nonumber R_i(\theta)& =\int_{0}^{\frac{b_i-b_{i,b}}{b_{i,b}}v} \frac{v+v_e}{d_l}D_i(\theta|v_e\geq0)f_{v_e}(v_e)dv_e \\
                     & +\int_{\frac{b_{i-1}-b_{i,e}}{b_{i,e}}v}^{0}  \frac{v+v_e}{d_l}D_i(\theta|v_e<0)f_{v_e}(v_e)dv_e
\end{align}
where $f_{v_e}(\cdot)$ represents the probability density function of $v_e$, and it is assumed to be normal. We perform some numerical results to investigate the effects of beam design parameters on the average data rate in the presence of estimation error. Required parameters for the numerical results are listed in Table 1.

\begin{table}
\centering
\caption{parameter used in our simulation results}
\renewcommand{\arraystretch}{1.5}
\begin{tabular}{|l|c|c|}\hline
\textbf{Parameter}	       & \textbf{Symbol}	& \textbf{value} 	\\ \hline
Carrier frequency	& $f_c$		& 60 GHz	\\ \hline
Path loss exponent	& $n$		& 2			\\ \hline
Effective Isotropically Radiated Power & \textit{EIRP} & 20 dBm\\ \hline
Covered distance by RSU				   & $d_l$	 & 100 m \\ \hline
see Fig.2					 			   & $d_0$	 & 3 m   \\ \hline
RSU height			& $H_R$		& 7 m		\\ \hline
Vehicle height	 	& $H_V$		& 1.5 m		\\ \hline
Vehicle speed	 	& $v$		& 25 m/s	\\ \hline
Noise figure	 	& $NF$		& 6 dB		\\ \hline
Bandwidth	 		& $B$		& 2.16 GHz	\\ \hline
Shadowing margin	& $w$		& 10 dB		\\ \hline
Lane width	 		& $W_l$		& 3.5 m		\\ \hline
\end{tabular}
\end{table}

\begin{figure}[t]
\centering
\includegraphics[width=0.99\linewidth]{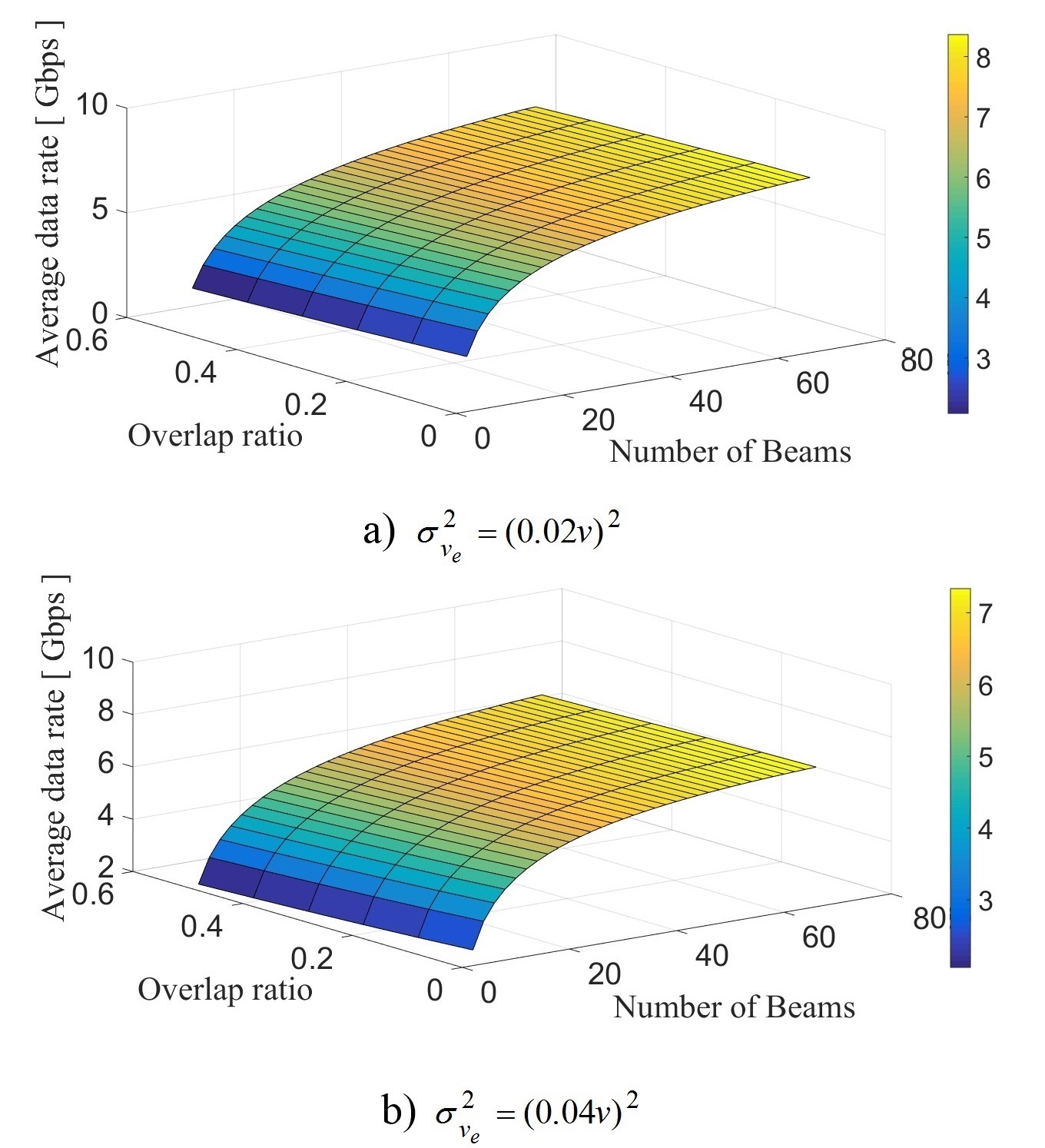}
\caption{Average data rates for equal beam, beam switching approach for different values of $v_e$}
\label{fig:data rate of equal Beam design}
\end{figure}

\begin{figure}[h]
\centering
\includegraphics[width=0.99\linewidth]{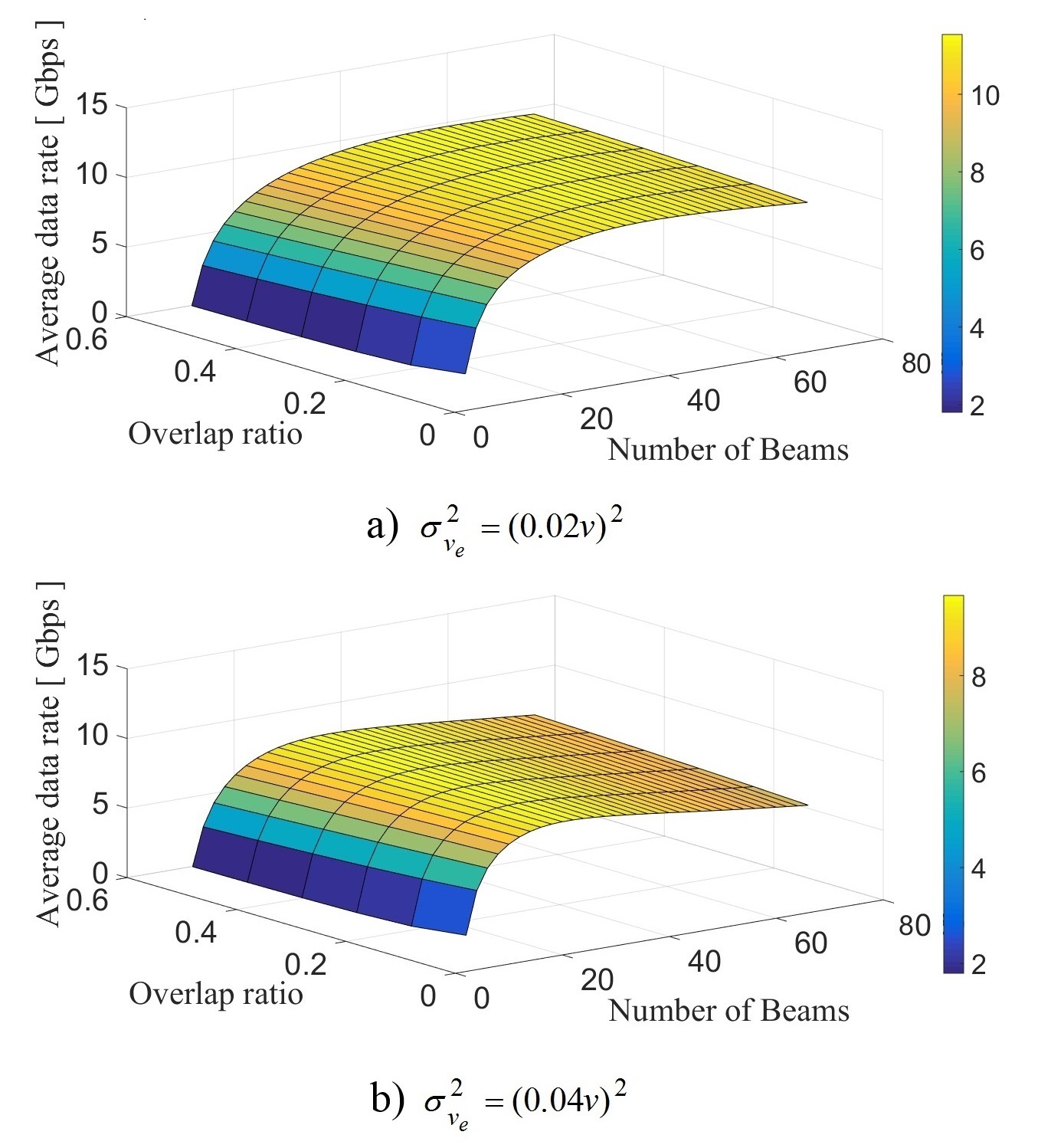}
\caption{Average data rates for equal coverage beam switching approach for different values of $v_e$}
\label{fig:data rate of equal coverage design}
\end{figure}

Three dimensional plots of the average data rate versus beam overlap ratio and the total number of beams ($N_b$), for the two beam design strategies of equal beamwidth and equal coverage are shown in Fig. \ref{fig:data rate of equal Beam design} and \ref{fig:data rate of equal coverage design}. We can see the effects of overlap ratio and $N_b$  in the presence of two speed estimation error variances of $(0.02v)^2$ and $(0.04v)^2$. It can be observed that the maximum achievable data rates at the case of equal coverage design are roughly 1.5 times higher than the equal beam approach, showing agreement with results in \cite{V.Va2016}. However, results for the equal coverage are more sensitive to the estimation error variance, and the equal beam design is more resistant to the error changes.

\subsection{Average Outage Time}
Having beam misalignment leads to an outage. This could be happening due to either position error, or speed error. It is assumed that for large values of position estimation error, RSU can solve the issue somehow by doing some correction of beam alignment to assign the right beam. We recall that the position and speed information are only once are provided at the first time the vehicle enters the coverage area of the RSU.
We focus on the case of smaller values of position error, and/or speed estimation error that cause outage while the vehicle passes the coverage area. Referring back to Fig. \ref{fig:RSU_beam_sectors} and separating negative and non-negative values of $v_e$, the outage time for the $i$-th beam can be expressed as \cite{V.Va2016}:
\begin{align}
\nonumber T_{out,i}(\theta|v_e\geq0)  =  \frac{b_{i+1,b}-b_{i,b}}{v}
Q(\frac{b_i-b_{i,b}}{b_{i,b}}\frac{v}{\sigma_v}) \\
+\int_{\frac{b_i-b_{i+1,b}}{b_{i+1,b}}v}^{\frac{b_i-b_{i,b}}{b_{i,b}}v}(\frac{b_{i+1,b}}{v}-\frac{b_i}{v+v_e})f_{v_e}(v_e)dv_e
\end{align}

\begin{align}
\nonumber T_{out,i}(\theta|v_e<0) = \frac{b_{i,e}-b_{i-1,e}}{v}Q(\frac{b_{i,e}-b_{i-1}}{b_{i,e}}\frac{v}{\sigma_v}) \\
+\int_{\frac{b_{i-1}-b_{i,e}}{b_{i,e}}v}^{\frac{b_i-b_{i,e}}{b_{i,e}}v}(\frac{b_{i}}{v+v_e}-\frac{b_{i,e}}{v})f_{v_e}(v_e)dv_e
\end{align}

In our simulation results, we demonstrate the \emph{outage percentage} as $T_{out,i}(\theta|v_e)$ divided by the total time that the vehicle covered by the RSU.

\begin{figure}[t]
\centering
\includegraphics[width=0.99\linewidth]{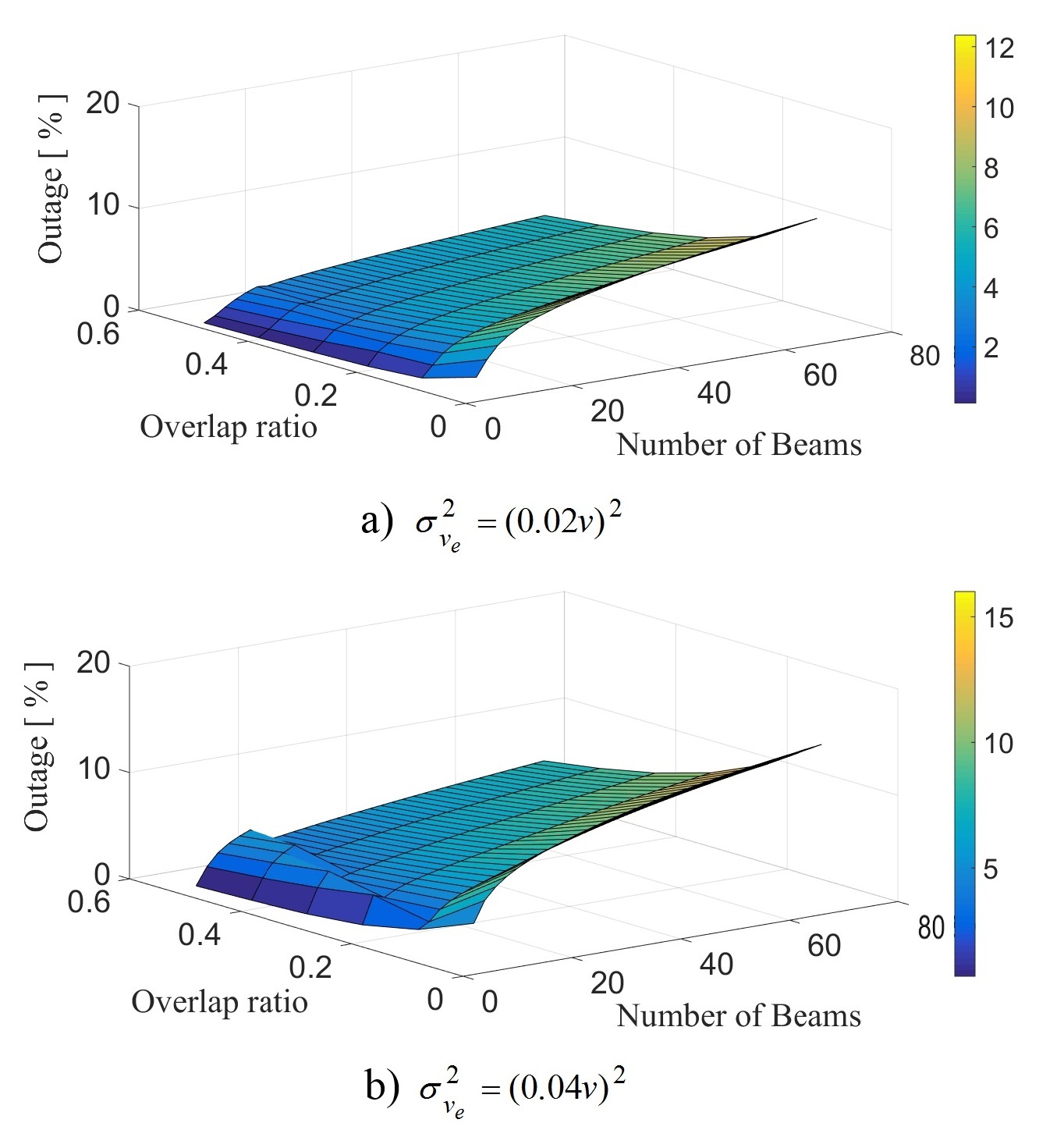}
\caption{Outage time (percentage) for equal beam, beam switching approach for different values of $v_e$}
\label{fig:outage of equal Beam design}
\end{figure}

\begin{figure}
\centering
\includegraphics[width=0.99\linewidth]{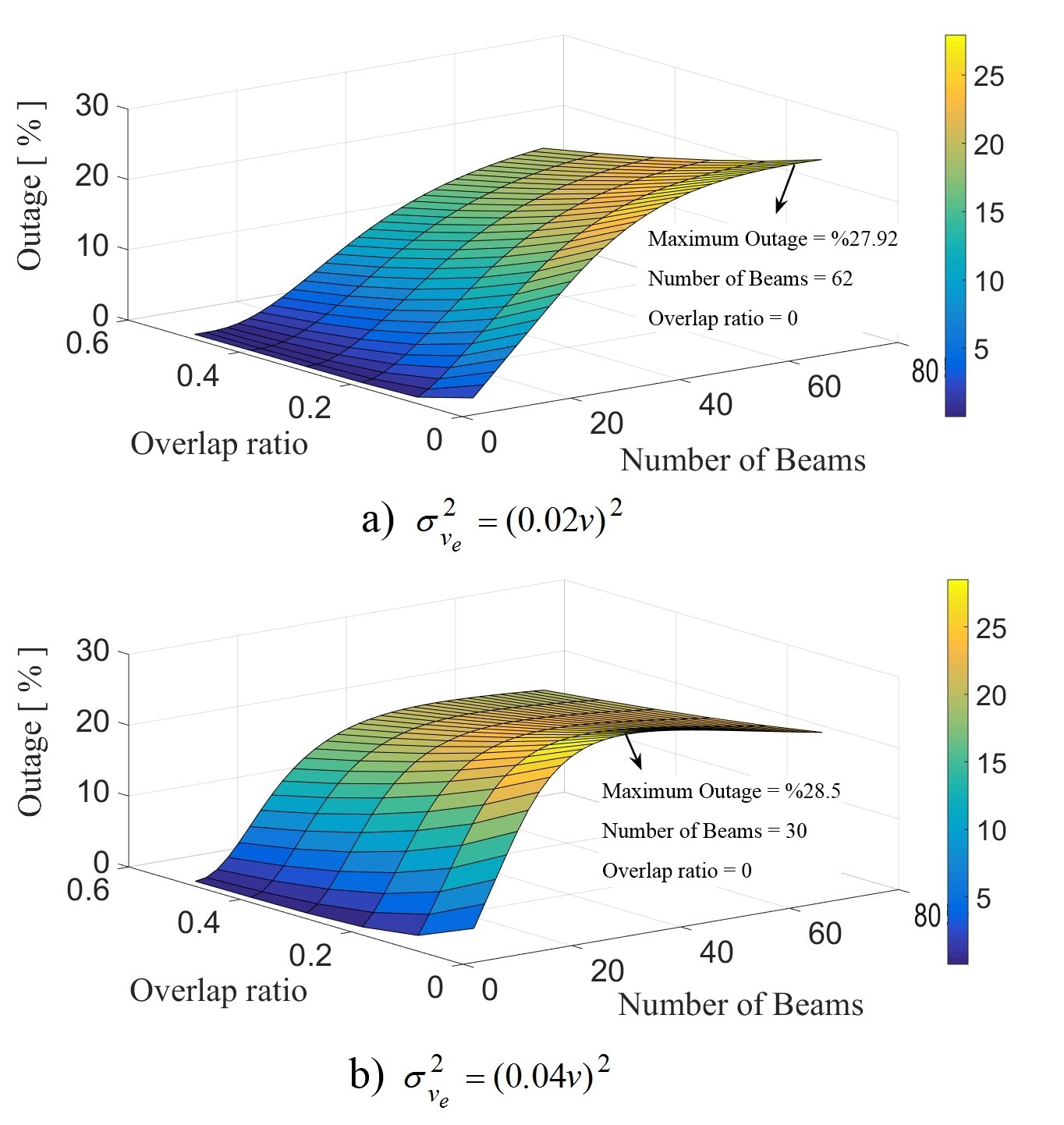}
\caption{Outage time (percentage) for equal coverage beam switching approach for different values of $v_e$}
\label{fig:outage of equal coverage design}
\end{figure}

Fig. \ref{fig:outage of equal Beam design} and \ref{fig:outage of equal coverage design} illustrates the outage time versus two design parameters of overlap ratio and the number of beams for equal coverage and equal beam designs. Despite the results of average data rates in Fig. \ref{fig:data rate of equal Beam design} and \ref{fig:data rate of equal coverage design}, we see that the equal beam presents lower outages compared to the equal coverage. Furthermore, the equal beam is less sensitive to the estimation error as seen in Fig. \ref{fig:data rate of equal Beam design} and \ref{fig:data rate of equal coverage design}.

We propose a new design efficiency metric in the following section, to compare the performances of two different designing approaches of equal beam and equal coverage in the presence of estimation error.

\section{Beam Design Efficiency}
As we see from Figures 3-6, increasing the number of beams ($N_b$) leads to increasing the average data rate. However, the average outage time is also increasing and it is not desired. Therefore, we propose a \emph{Beam Design Efficiency} (BDE) function to have a tradeoff between data rate and outage time as follows:
\begin{equation}
\textit{BeamDesignEfficiency} = \alpha\times \textit DataRate- \beta \times \textit Outage
\end{equation}
where $\alpha$ and $\beta$ are weighting coefficients that are determined from the following equations:
\begin{equation}
\alpha \times \max(\textit DataRate)-\beta \times \min(\textit Outage)=1
\end{equation}
\begin{equation}
\alpha \times \min(\textit DataRate)-\beta \times \max(\textit Outage)=0
\end{equation}

We note that the maximum and minimum values of Average data rate, and outage time are calculated while the following conditions are met:
\begin{itemize}
\item $\theta_1+\theta_2+...+\theta_{N_b}=\theta_{RSU}  \quad \forall \theta_i>0, \quad i=1,2,...,N_b$
\item Each beam has a maximum of $\%50$ overlap with its adjacent beams.
\end{itemize}

\begin{figure}[h]
\centering
\includegraphics[width=0.99\linewidth]{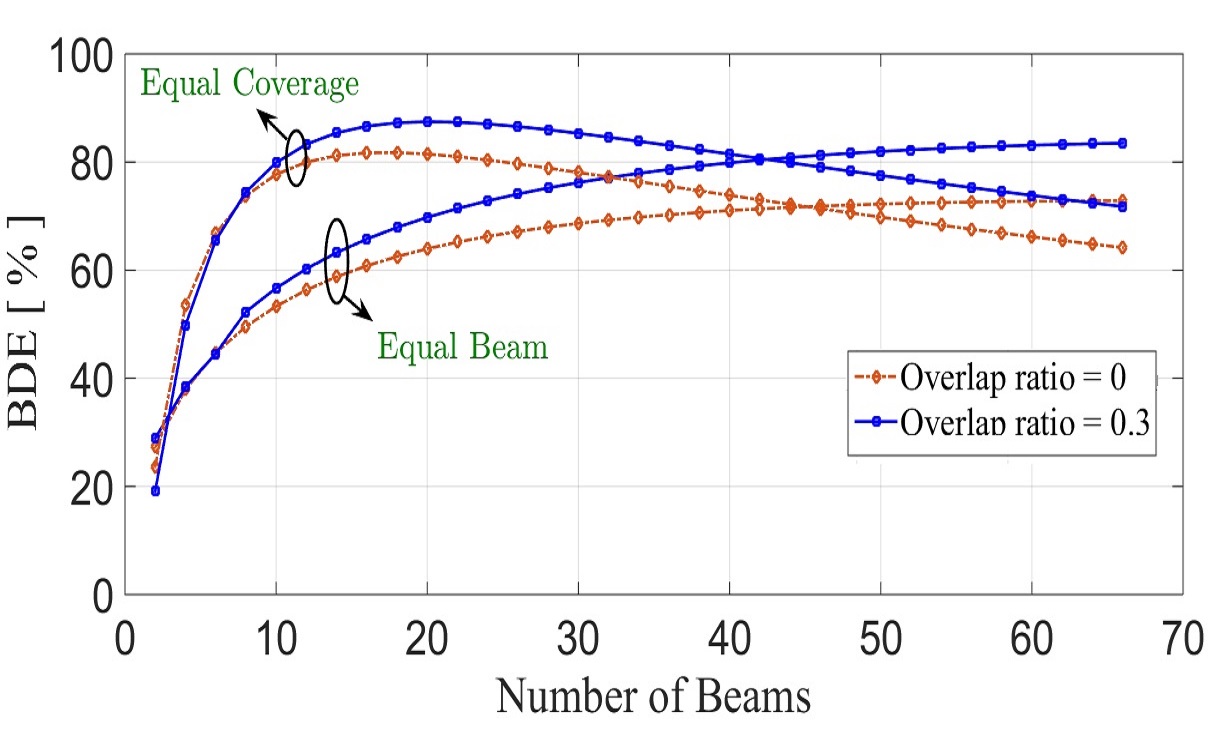}
\caption{Beam design efficiency for $\sigma_{v_e}^2=(0.04v)^2$}
\label{fig:Beam design efficiency}
\end{figure}

According to Fig. \ref{fig:RSU_beam_sectors}, $\theta_{RSU}$ can be calculated as follows:
\begin{equation}
\theta_{RSU}=2\tan^{-1}(\frac{d_l}{2d_0})
\end{equation}
Fig. \ref{fig:Beam design efficiency} shows the beam design efficiency for the equal beam and equal coverage versus the number of beams, having an estimation error variance of $\sigma_{v_e}^2=(0.04v)^2$. Two values of zero and 30 percent are considered for the overlap ratio between beams.
Having the number of beams below 41, we observe that the equal coverage method outperforms the equal beam design. As we increase $N_b$  above 42 or 43, the equal coverage is less efficient than the equal beam. This would be interpreted as a stronger effect of the outage time causing the equal coverage performance falls down.

\section{Conclusion}
In this paper, beam switching techniques are addressed for V2I communications, that use initial speed and position estimation of a vehicle entering the coverage area of an RSU, to reduce beam alignment overheads. However, the speed and/or (small) position estimation error of the vehicles lead to beam misalignment problem. We studied the effects of the number of beams (or equivalently the beamwidth of beams), and overlapping between beams, on the system performance, as available design parameters to combat this effect. We considered two beam design strategies for beam switching at the RSU: having equal coverage area, or equal beamwidth for all beams. However, no solution existed in the literature to answer that how these parameters should be selected? Here, we proposed a beam design efficiency criterion that allows us to select the right values of design parameters and decide which beam design approach is better in the presence of estimation error.

\bibliographystyle{IEEEtran}

\end{document}